\documentclass{article}
\usepackage[T2A]{fontenc}
\usepackage[cp1251]{inputenc}
\usepackage[russian]{babel}
\usepackage{graphicx}
\begin{document}


\vskip 4mm

\centerline{\large \bf Effect of nonmagnetic and magnetic
impurities}
\centerline{\large \bf on the specific heat jump in anisotropic
superconductors}

\vskip 2mm

\centerline{Leonid A. Openov}

\vskip 2mm

\centerline{\it Moscow Engineering Physics Institute (State University)}
\centerline{\it 115409 Moscow, Russia}
\centerline{opn@supercon.mephi.ru}

\vskip 4mm

\begin{quotation}

The specific-heat jump $\Delta C$ at a critical temperature $T_c$ in an
anisotropic superconductor containing both potential and spin-flip scatterers
is calculated within a weak-coupling mean-field approximation. It
is shown that the presence of even a small amount of spin-flip scatterers in
the sample leads to a drastic change in the dependence of $\Delta C$ on
$T_c$ in a disordered $(d+s)$-wave or a strongly anisotropic $s$-wave
superconductor. The implications for experimental tests for the presence of
an $s$-wave admixture in the superconducting order parameter of high-$T_c$
superconductors are discussed.

\end{quotation}

\vskip 6mm

\centerline{\bf I. INTRODUCTION}

\vskip 2mm

Although there is a considerable evidence in favour of predominantly $d$-wave
in-plane symmetry of the superconducting order parameter $\Delta({\bf p})$ in
high-temperature superconductors (HTSCs) \cite{Annett}, a number of
experiments point to an admixture of the $s$-wave component to $d$-wave
$\Delta({\bf p})$, implying a mixture of $d$-wave and $s$-wave components,
see references in Ref. \cite{Annett}, or even to a strongly anisotropic
$s$-wave $\Delta({\bf p})$, see Refs. \cite{Zhao,Brandow}. The non-pure
$d$-wave symmetry of $\Delta({\bf p})$ is supported in part by a
long tail suppression of the critical temperature, $T_c$, by defects and
impurities \cite{Valles,Sun} characteristic for a nonzero value of the Fermi
surface (FS) average $\langle\Delta({\bf p})\rangle_{FS}$, while
$\langle\Delta({\bf p})\rangle_{FS}=0$ for a $d$-wave pairing. However,
neither the presence nor the absence of an isotropic $s$-wave component of
$\Delta({\bf p})$ in HTSCs has not been demonstrated unequivocally yet.

Recently Hara\'n {\it et al.} \cite{Haran} have shown that significant
features attributed to the $s$-wave part of $\Delta({\bf p})$ may be seen in
the measurements of the specific heat jump $\Delta C$ at $T_c$ in disordered
$(d+s)$-wave superconductors with nonmagnetic impurities. It was found in
Ref. \cite{Haran} that there is a qualitative difference in the evolution of
$\Delta C$ under disordering in $d$-wave and $(d+s)$-wave superconductors.
While in a $d$-wave superconductor, the value of $\Delta C$ normalized by the
normal state specific heat $C_n(T_c)$ monotonously goes to zero as $T_c$ is
suppressed by nonmagnetic disorder, in a $(d+s)$-wave superconductor there is
a minimum at the curve of $\Delta C/C_n(T_c)$ versus $T_c$. The location of
this minimum depends on a specific weight of an isotropic $s$-wave component
in the $(d+s)$-wave $\Delta({\bf p})$, i. e., on a specific value of
$\langle\Delta({\bf p})\rangle_{FS}$. It was suggested in Ref. \cite{Haran}
that this effect may be used as a test for the presence of an $s$-wave
admixture in HTSCs.

The authors of Ref. \cite{Haran} restricted their consideration to the case
of nonmagnetic disorder (i. e., potential scattering of charge carriers) only.
Note, however, that a lot of experiments give evidence for the presence of
spin-flip scatterers (along with potential ones) in non-stoichiometric
HTSCs, e. g., in oxygen-deficient, doped or irradiated samples (see, e. g.,
the references in Ref. \cite{Openov}). The spin-flip scattering of charge
carriers by magnetic impurities (even though they may be present in small
proportion) can strongly modify the value of $T_c$ as well as other
characteristics of a superconductor in the vicinity of $T_c$
\cite{Openov,Openov2}. It is therefore instructive to elucidate if an account
for spin-flip scattering will change the conclusion drawn in
Ref. \cite{Haran} concerning the impurity effect on $\Delta C$ in anisotropic
superconductors.

The influence of the ratio of spin-flip to non spin-flip scattering rates
on the specific heat jump in a weakly anisotropic superconductor has been
considered earlier by Okabe and Nagi \cite{Okabe}. In the present paper we
study the combined effect of potential and spin-flip scatterers on
$\Delta C$ in a superconductor with arbitrary anisotropy of
$\Delta({\bf p})$, including $s$-wave, $d$-wave and $(d+s)$-wave as
particular cases. In what follows, we make use of the following
approximations: (i) we consider the pairing interactions in the weak-coupling
limit of the BCS model and assume that the pairing potential can be taken in
a factorizable form; (ii) we treat the impurity scattering in the Born limit;
(iii) we assume an $s$-wave scattering of charge carriers by both potential
and spin-flip scatterers, so that the scattering matrix elements are momentum
independent; (iv) we neglect any dynamic pair breaking effects
\cite{Williams}.

One should keep in mind that the mean-field BCS theory does
not describe the effects of spatial variation of the order parameter in the
vicinity of impurities \cite{Franz}. These effects may be significant in
superconductors with short coherence length. However, experiments on the
impurity induced suppression of $T_c$ in anisotropic superconductors are, in
fact, rather well described within the mean-field Abrikosov-Gor'kov approach
\cite{Abrikosov} (see, e. g., \cite{Dalichaouch}). So, the spatial variation
of the order parameter may appear to have a little influence on the physical
characteristics, including $\Delta C$, in the vicinity of $T_c$.
Note also that strong-coupling corrections \cite{Carbotte}
result just in renormalization of the scattering rates
\cite{Radtke} and overall increase in the magnitude of $\Delta C$, without
qualitative changes in the dependence of $\Delta C$ versus $T_c$
\cite{Prohammer,El-Hagary}. As for the use of the weak scattering (Born)
limit, it seems to be justified by a very close similarity of $\Delta C$
versus $T_c$ curves calculated in the Born and unitary limits for the case of
purely nonmagnetic disorder in $d$-wave and $(d+s)$-wave superconductors
\cite{Haran,Prohammer,Puchkaryov}. We set $\hbar=k_B=1$ throughout the paper.

\vskip 6mm

\centerline{\bf II. FORMALISM}

\vskip 2mm

The specific heat jump is defined as $\Delta C = C_s(T_c)-C_n(T_c)$, where
the subscripts $s$ and $n$ refer to the superconducting and normal state
respectively, $C_n(T_c)=(2\pi^2/3)N(0)T_c$, and $N(0)$ is the density of
electron states per spin at the Fermi level. The value of $\Delta C$ can be
expressed in terms of the thermodynamic potential $\Omega$ as
\begin{equation}
\Delta C=
-T_c\left[\frac{\partial^2(\Omega_s-\Omega_n)}{\partial T^2}\right]_{T=T_c}.
\label{DeltaC}
\end{equation}
For the value of the difference $\Omega_s-\Omega_n$ one has (see, e. g., Ref.
\cite{Abrikosov})
\begin{equation}
\Omega_s-\Omega_n=\int_0^{\Delta^2}\frac{d(V_0^{-1})}
{d\Delta^2}\Delta^2d\Delta^2,
\label{deltaOmega}
\end{equation}
where $V_0$ is the pairing energy that determines the magnitude of the
phenomenological factorizable pairing potential of the form
$V({\bf p},{\bf p}^{\prime})=-V_0\phi({\bf n})\phi({\bf n}^{\prime})$,
$\Delta$ is the amplitude of the superconducting order parameter
$\Delta({\bf p})=\Delta\phi({\bf n})$,
${\bf n}={{\bf p}}/p$ is a unit vector along the momentum, and the function
$\phi({\bf n})$ specifies the symmetry and anisotropy of $\Delta ({\bf p})$
in the momentum space [e. g., $\phi({\bf n})\equiv$ const for isotropic
$s$-wave pairing; $\phi({\bf n})=\cos (2\varphi)$ for a specific case of
$d$-wave pairing, where $\varphi$ is an angle between the vector ${\bf n}$
and the $x$-axis; $\phi({\bf n})=r+\cos (2\varphi)$ for a specific case of
mixed $(d+s)$-wave pairing, the constant $r$ being the measure of the partial
weight of $s$-wave component in $\Delta({\bf p})]$.

To find the dependence of $V_0^{-1}$ on $\Delta$, we make use of the set of
mean-filed self-consistent equations for a superconductor containing both
nonmagnetic and magnetic impurities (see Ref. \cite{Openov}):
\begin{equation}
\Delta({\bf p})=-T\sum_{\omega}\sum_{{\bf k}}
V({\bf p},{\bf k})\frac{\Delta_\omega({\bf k})}
{\omega^{\prime 2}+\xi^2({\bf k})+|\Delta_\omega({\bf k})|^2}~,
\label{Delta1}
\end{equation}
\begin{equation}
\Delta_\omega({\bf p})=\Delta({\bf p})+
(c_n|u_n|^2+c_m|u_m^{pot}|^2-c_m|u_m^{ex}|^2)\sum_{{\bf k}}
\frac{\Delta_\omega({\bf k})}{\omega^{\prime 2}+\xi^2({\bf k})+
|\Delta_\omega({\bf k})|^2}~,
\label{Delta_omega}
\end{equation}
\begin{equation}
\omega^{\prime}=\omega-i(c_n|u_n|^2+c_m|u_m^{pot}|^2+c_m|u_m^{ex}|^2)
\sum_{{\bf k}}\frac{i\omega^{\prime}+\xi({\bf k})}
{\omega^{\prime 2}+\xi^2({\bf k})+|\Delta_\omega({\bf k})|^2}~,
\label{omega'}
\end{equation}
where $T$ is the temperature; $\xi({\bf p})$ is the quasiparticle energy
measured from the chemical potential; $\omega=\pi T(2n+1)$ are Matsubara
frequencies; $c_n$ and $c_m$ are the concentrations of nonmagnetic and
magnetic impurities, respectively; $u_n$ is the matrix element for potential
electron scattering by an isolated nonmagnetic impurity; $u_m^{pot}$ and
$u_m^{ex}$ are the matrix elements for, respectively, potential and exchange
(spin-flip) scattering by an isolated magnetic impurity.

Restricting all electron momenta in Eqs. (\ref{Delta1})-(\ref{omega'})
to the FS, replacing $\sum_{{\bf k}}$ by $N(0)\int d\xi({\bf k})
\int_{FS}d\Omega_{\bf k}/|\partial\xi({\bf k})/\partial{\bf k}|$, and
integrating over $\xi({\bf k})$, one has
\begin{equation}
V_0^{-1}=\pi N(0)\langle\phi^2({\bf n})\rangle_{FS} T
\sum_{\omega}f(\omega,\Delta)~,
\label{V0^-1}
\end{equation}
\begin{equation}
\Delta_\omega({\bf p})=\Delta({\bf p})+
\frac{1}{2}\left (\frac{1}{\tau_p}-\frac{1}{\tau_s}\right )
\left < \frac{\Delta_\omega({\bf p})}
{\sqrt{\omega^{\prime 2}+|\Delta_\omega({\bf p})|^2}} \right >_{FS}~,
\label{Delta_omega1}
\end{equation}
\begin{equation}
\omega^{\prime}=\omega+
\frac{1}{2}\left (\frac{1}{\tau_p}+\frac{1}{\tau_s}\right )
\omega^{\prime}\left < \frac{1}
{\sqrt{\omega^{\prime 2}+|\Delta_\omega({\bf p})|^2}} \right >_{FS}~,
\label{omega'1}
\end{equation}
where
\begin{equation}
f(\omega,\Delta)=\frac{1}{\langle\phi^2({\bf n})\rangle_{FS}\Delta}
\left < \frac{\phi({\bf n})\Delta_\omega({\bf p})}
{\sqrt{\omega^{\prime 2}+|\Delta_\omega({\bf p})|^2}} \right >_{FS}~,
\label{f}
\end{equation}
the angular brackets $\langle ... \rangle_{FS}$ stand for a FS average,
\begin{equation}
\langle ... \rangle_{FS}=\int_{FS}(...)\frac{d\Omega_{\bf p}}{|\partial
\xi({\bf p})/\partial {\bf p}|}\Biggl/\int_{FS}\frac{d\Omega_
{\bf p}}{|\partial\xi({\bf p})/\partial {\bf p}|}~,
\label{average}
\end{equation}
and we have introduced the electron relaxation times $\tau_p$ and $\tau_s$
for potential and spin-flip scattering, respectively:
\begin{equation}
\frac{1}{\tau_p}=2\pi (c_n|u_n|^2+c_m|u_m^{pot}|^2)N(0),~
\frac{1}{\tau_s}=2\pi c_m|u_m^{ex}|^2N(0).
\label{tau}
\end{equation}

\vskip 6mm

\centerline{\bf III. RESULTS}

\vskip 2mm

Making use of a standard procedure, one can transform Eq. (\ref{V0^-1}) into
\begin{equation}
\ln\left(\frac{T}{T_{c0}}\right)=\pi T\sum_{\omega}
\left(f(\omega,\Delta)-\frac{1}{|\omega|}\right)~,
\label{Tc0}
\end{equation}
where $T_{c0}$ is the value of $T_c$ in the absence of impurities. At
$\Delta=0$, one obtains from Eqs. (\ref{Delta_omega1})-(\ref{f}), (\ref{Tc0})
an expression for $T_c$ as a function of potential and spin-flip scattering
rates, $\rho_p=1/4\pi\tau_pT_c$ and
$\rho_s=1/2\pi\tau_sT_c$, respectively (see Ref. \cite{Openov})
\begin{equation}
\ln\left(\frac{T_{c0}}{T_c}\right)=
\frac{\langle\phi({\bf n})\rangle^2_{FS}}{\langle\phi^2({\bf n})\rangle_{FS}}
\left[\Psi\left(\frac{1}{2}+\rho_s\right)-\Psi\left(\frac{1}{2}\right)\right]
+\left(1-
\frac{\langle\phi({\bf n})\rangle^2_{FS}}{\langle\phi^2({\bf n})\rangle_{FS}}
\right)\left[\Psi\left(\frac{1}{2}+\rho\right)-
\Psi\left(\frac{1}{2}\right)\right]~,
\label{Tc}
\end{equation}
where $\rho=\rho_p+\rho_s/2$ is the total scattering rate and $\Psi$ is the
digamma function.

Expanding $f(\omega,\Delta)$ in powers of $\Delta^2$ up to $\Delta^2$ and
differentiating Eq. (\ref{V0^-1}) with respect to $\Delta^2$, one has from
Eq. (\ref{deltaOmega}) in the vicinity of $T_c$
\begin{equation}
\Omega_s-\Omega_n=\frac{\pi}{2}N(0)\langle\phi^2({\bf n})\rangle_{FS}
\Delta^4(T)T\sum_{\omega}\left(\frac{df(\omega,\Delta)}{d\Delta^2}\right)
_{\Delta=0}.
\label{deltaOmega1}
\end{equation}
Next, taking Eq. (\ref{Tc}) into account, one has from Eq. (\ref{Tc0}) an
expression for $\Delta^2(T)$ in the vicinity of $T_c$
\begin{equation}
\Delta^2(T)=\left(\frac{T}{T_c}-1\right)\frac{
\langle\phi^2({\bf n})\rangle_{FS}-\rho\left[
\langle\phi^2({\bf n})\rangle_{FS}-\langle\phi({\bf n})\rangle^2_{FS}
\right]\Psi^{(1)}\left(\frac{1}{2}+\rho\right)-\rho_s
\langle\phi({\bf n})\rangle^2_{FS}\Psi^{(1)}\left(\frac{1}{2}+\rho_s\right)}
{\pi\langle\phi^2({\bf n})\rangle_{FS}
T_c\sum_{\omega}\left(\frac{df(\omega,\Delta)}{d\Delta^2}\right)_{\Delta=0}}.
\label{Delta^2}
\end{equation}
Here and below $\psi^{(n)}(z)$ are the polygamma functions [the n-th
derivatives of the digamma function $\psi(z)$] defined as
$\psi^{(n)}(z)=(-1)^{n+1}n!\sum_{k=0}^{\infty}(k+z)^{-(n+1)}$.

Substituting Eq. (\ref{Delta^2}) into Eq. (\ref{deltaOmega1}), we obtain from
Eq. (\ref{DeltaC}) the expression for the specific heat jump $\Delta C$
normalized by the specific heat in the normal state,
\begin{equation}
\frac{\Delta C}{C_n(T_c)}=-12
\frac{\biggl[\langle\phi^2({\bf n})\rangle_{FS}-
\rho\left[\langle\phi^2({\bf n})\rangle_{FS}-
\langle\phi({\bf n})\rangle^2_{FS}\right]
\Psi^{(1)}\left(\frac{1}{2}+\rho\right)-\rho_s
\langle\phi({\bf n})\rangle^2_{FS}
\Psi^{(1)}\left(\frac{1}{2}+\rho_s\right)\biggr]^2}
{(2\pi T_c)^3\langle\phi^2({\bf n})\rangle_{FS}
\sum_{\omega}\left(\frac{df(\omega,\Delta)}{d\Delta^2}\right)
_{\Delta=0}}.
\label{DeltaC1}
\end{equation}
Finally, after simple but rather cumbersome calculations, we find from
Eqs. (\ref{Delta_omega1})-(\ref{f}) an expression for the denominator in
Eq. (\ref{DeltaC1})
\begin{eqnarray}
&&(2\pi T_c)^3\langle\phi^2({\bf n})\rangle_{FS}
\sum_{\omega}\left(\frac{df(\omega,\Delta)}{d\Delta^2}\right)_{\Delta=0}
\nonumber \\
&&=\frac{\rho}{6}\Psi^{(3)}\left(\frac{1}{2}+\rho\right)\left[
\langle\phi^2({\bf n})\rangle_{FS}-
\langle\phi({\bf n})\rangle^2_{FS}\right]^2+
\frac{\rho_s}{6}\Psi^{(3)}\left(\frac{1}{2}+\rho_s\right)
\langle\phi({\bf n})\rangle^4_{FS}+
\frac{1}{2}\Psi^{(2)}\left(\frac{1}{2}+\rho_s\right)
\langle\phi({\bf n})\rangle^4_{FS}
\nonumber \\
&&+\frac{1}{2}\Psi^{(2)}\left(\frac{1}{2}+\rho\right)
\left[\langle\phi^4({\bf n})\rangle_{FS}-
4\langle\phi^3({\bf n})\rangle_{FS}\langle\phi({\bf n})\rangle_{FS}+
6\langle\phi^2({\bf n})\rangle_{FS}\langle\phi({\bf n})\rangle^2_{FS}-
3\langle\phi({\bf n})\rangle^4_{FS}\right]
\nonumber \\
&&+\frac{1}{\rho-\rho_s}\Psi^{(1)}\left(\frac{1}{2}+\rho\right)\left[
4\langle\phi^3({\bf n})\rangle_{FS}\langle\phi({\bf n})\rangle_{FS}-
11\langle\phi^2({\bf n})\rangle_{FS}\langle\phi({\bf n})\rangle^2_{FS}+
7\langle\phi({\bf n})\rangle^4_{FS}\right]
\nonumber \\
&&-\frac{5}{\rho-\rho_s}\Psi^{(1)}\left(\frac{1}{2}+\rho_s\right)
\langle\phi({\bf n})\rangle^2_{FS}\left[\langle\phi^2({\bf n})\rangle_{FS}-
\langle\phi({\bf n})\rangle^2_{FS}\right]
\nonumber \\
&&+\frac{\rho+\rho_s}{(\rho-\rho_s)^2}
\left[\Psi^{(1)}\left(\frac{1}{2}+\rho\right)+
\Psi^{(1)}\left(\frac{1}{2}+\rho_s\right)\right]
\langle\phi({\bf n})\rangle^2_{FS}\left[\langle\phi^2({\bf n})\rangle_{FS}-
\langle\phi({\bf n})\rangle^2_{FS}\right]
\nonumber \\
&&-\frac{4}{(\rho-\rho_s)^2}\left[
\Psi\left(\frac{1}{2}+\rho\right)-\Psi\left(\frac{1}{2}+\rho_s\right)\right]
\left[\langle\phi^3({\bf n})\rangle_{FS}\langle\phi({\bf n})\rangle_{FS}-
4\langle\phi^2({\bf n})\rangle_{FS}\langle\phi({\bf n})\rangle^2_{FS}+
3\langle\phi({\bf n})\rangle^4_{FS}\right]
\nonumber \\
&&-2\frac{\rho+\rho_s}{(\rho-\rho_s)^3}\left[
\Psi\left(\frac{1}{2}+\rho\right)-\Psi\left(\frac{1}{2}+\rho_s\right)\right]
\langle\phi({\bf n})\rangle^2_{FS}\left[\langle\phi^2({\bf n})\rangle_{FS}-
\langle\phi({\bf n})\rangle^2_{FS}\right].
\label{df/dDelta^2}
\end{eqnarray}

Eqs. (\ref{DeltaC1}) and (\ref{df/dDelta^2}), together with Eq. (\ref{Tc})
for $T_c$, give the value of the specific heat jump for a superconductor that
is characterized by an arbitrary anisotropy of $\Delta({\bf p})$ [i. e., by
an arbitrary angular function $\phi({\bf n})$] and contains, in general, both
potential and spin-flip scatterers. In particular cases of (i) spin-flip
scattering in an isotropic $s$-wave superconductor
with $\phi({\bf n})\equiv$ const and (ii)
potential scattering in a highly anisotropic (e. g., $d$-wave)
superconductor with $\langle\phi({\bf n})\rangle_{FS}=0$, Eqs.
(\ref{DeltaC1}) and (\ref{df/dDelta^2}) reduce to the well-known expressions
\cite {Haran,Puchkaryov,Skalski,Hirschfeld,Suzumura}
\begin{equation}
\frac{\Delta C}{C_n(T_c)}=-12
\frac{\left[1-\rho_s\Psi^{(1)}\left(\frac{1}{2}+\rho_s\right)\right]^2}
{\frac{1}{2}\Psi^{(2)}\left(\frac{1}{2}+\rho_s\right)+
\frac{\rho_s}{6}\Psi^{(3)}\left(\frac{1}{2}+\rho_s\right)}
\label{Smagnetic}
\end{equation}
and
\begin{equation}
\frac{\Delta C}{C_n(T_c)}=-12
\frac{\left[1-\rho\Psi^{(1)}\left(\frac{1}{2}+\rho\right)\right]^2}
{\frac{1}{2}\frac{\langle\phi^4({\bf n})\rangle_{FS}}
{\langle\phi^2({\bf n})\rangle_{FS}^2}
\Psi^{(2)}\left(\frac{1}{2}+\rho\right)+
\frac{\rho}{6}\Psi^{(3)}\left(\frac{1}{2}+\rho\right)},
\label{Dnonmagnetic}
\end{equation}
respectively \cite{Notes}. In the case of a superconductor that has an
arbitrary anisotropy of $\Delta({\bf p})$ but contains nonmagnetic impurities
only, Eqs. (\ref{DeltaC1}) and (\ref{df/dDelta^2}) reduce to the results of
Hara\'n {\it et al.} \cite{Haran,Notes}. In the absence of any impurities
($T_c=T_{c0}$), one has
\begin{equation}
\frac{\Delta C}{C_n(T_{c0})}=\frac{12}{7\zeta(3)}
\frac{\langle\phi^2({\bf n})\rangle^2_{FS}}
{\langle\phi^4({\bf n})\rangle_{FS}},
\label{DeltaC0}
\end{equation}
where $\zeta(3)\approx$ 1.202 is the Riemann zeta function.
Eq. (\ref{DeltaC0}) has been widely used to analyze the effect of anisotropy
of $\Delta({\bf p})$ on the specific heat jump in clean superconductors (see,
e. g., \cite{LANL}). For a clean isotropic [$\phi({\bf n})\equiv$ const]
superconductor we arrive at a familiar BCS result,
$\Delta C/C_n(T_{c0})=12/7\zeta(3)\approx$ 1.426.

\vskip 6mm

\centerline{\bf IV. DISCUSSION}

\vskip 2mm

In what follows, we shall model the dependence of
$\Delta({\bf p})=\Delta\phi({\bf n})$ by the angular function
$\phi({\bf n})=r+\cos (2\varphi)$. The value of $r=0$ corresponds to $d$-wave
pairing, while $r\rightarrow \infty$ ($\Delta r\rightarrow$const)
in an isotropic $s$-wave superconductor. The smaller is the value of $r$,
the higher is the anisotropy of $\Delta ({\bf p})$. The moments
$\langle\phi({\bf n})\rangle_{FS}$, $\langle\phi^2({\bf n})\rangle_{FS}$,
$\langle\phi^3({\bf n})\rangle_{FS}$, and
$\langle\phi^4({\bf n})\rangle_{FS}$ that enter Eqs. (\ref{Tc}),
(\ref{DeltaC1}), and (\ref{df/dDelta^2}) are equal to $r$, $r^2+1/2$,
$r^3+3r/2$, and $r^4+3r^2+3/8$, respectively.

Note that the value of the specific heat jump in a clean superconductor is a
nonmonotonous function of $r$. It follows from Eq. (\ref{DeltaC0}) that the
normalized specific heat jump initially decreases with $r$ from
$\Delta C/C_n(T_{c0})\approx$ 0.951 at $r=0$ down to
$\Delta C/C_n(T_{c0})\approx$ 0.666 at $r=\sqrt{3/8}\approx 0.612$ and next
increases again up to $\Delta C/C_n(T_{c0})=12/7\zeta(3)\approx$ 1.426
at $r\rightarrow\infty$, see Fig. 1.

Now let us analyze the behavior of $\Delta C/C_n(T_c)$ versus $T_c$ upon
addition of magnetic and/or nonmagnetic impurities to the initially clean
sample with the critical temperature $T_{c0}$. We note that nonmagnetic
impurities result in the potential scattering only, while magnetic impurities
generally result in both spin-flip and potential scattering. In this respect,
the combined effect of nonmagnetic and magnetic impurities has much in common
with the effect of magnetic impurities only, the difference being in the
ratio of potential to spin-flip scattering rates as a function of impurity
concentrations.

\vskip 6mm

\centerline{\bf A. Nonmagnetic disorder}

\vskip 2mm

First, we consider the case
that there are no magnetic impurities in the sample, i. e., $\rho_s=0$ and,
hence, $\rho=\rho_p$ (see also Ref. \cite{Haran}).
At low concentration of nonmagnetic impurities,
i. e., at $(T_{c0}-T_c)/T_{c0}<<1$, one has from Eqs. (\ref{DeltaC1}) and
(\ref{df/dDelta^2})
\begin{equation}
\frac{\Delta C}{C_n(T_c)}=\frac{12}{7\zeta(3)}
\frac{\langle\phi^2({\bf n})\rangle^2_{FS}}
{\langle\phi^4({\bf n})\rangle_{FS}}\left[
1-2\frac{T_{c0}-T_c}{T_{c0}}\left(1+\frac{\pi^2}{42\zeta(3)}\cdot
\frac{\langle\phi^2({\bf n})\rangle_{FS}}{\langle\phi^4({\bf n})\rangle_{FS}}
\cdot
\frac{4\langle\phi^3({\bf n})\rangle_{FS}\langle\phi({\bf n})\rangle_{FS}-
3\langle\phi^4({\bf n})\rangle_{FS}-\langle\phi^2({\bf n})\rangle^2_{FS}}
{\langle\phi^2({\bf n})\rangle_{FS}-\langle\phi({\bf n})\rangle^2_{FS}}
\right)\right]~,
\label{DeltaC2}
\end{equation}
where we took into account that
$\rho=(2/\pi^2)(1-T_c/T_{c0})\langle\phi^2({\bf n})\rangle_{FS}/
[\langle\phi^2({\bf n})\rangle_{FS}-\langle\phi({\bf n})\rangle^2_{FS}]$ at
$(T_{c0}-T_c)/T_{c0}<<1$, see Eq. (\ref{Tc}) and Ref. \cite{Openov}. Note
that the term in round brackets in Eq. (\ref{DeltaC2}) changes sign from
positive to negative as $r$ increases up to $r_0\approx 1.75$.
Analysis of Eqs. (\ref{DeltaC1}), (\ref{df/dDelta^2}), and
(\ref{DeltaC2}) shows that in weakly anisotropic
superconductors with $r>r_0$, the normalized specific
heat jump increases monotonously up to $\Delta C/C_n(T_c)=12/7\zeta(3)$ as
$T_c$ is suppressed by nonmagnetic impurities, in a close agreement with the
behaviour of $\Delta C/C_n(T_c)$ in two-gap superconductors whose
thermodynamics is, to some respect, similar to that of anisotropic
superconductors, see Ref. \cite{Pokrovsky} (note that in an isotopic
superconductor, $r\rightarrow\infty$, the nonmagnetic disorder has no effect
on both $T_c$ and $\Delta C/C_n(T_c)$, see Refs. \cite{Abrikosov,Anderson}).
Contrary, at $0<r<r_0$, the normalized
specific heat jump initially decreases with
decreasing $T_c$, passes through a minimum at $T_c/T_{c0}=(T_c/T_{c0})^*$,
and then increases up to $\Delta C/C_n(T_c)=12/7\zeta(3)$ as
$T_c\rightarrow 0$. The position of the minimum at the curve of
$\Delta C/C_n(T_c)$ versus $T_c/T_{c0}$ depends on $r$, the value of
$(T_c/T_{c0})^*$ being reduced from 1 to 0 as $r$ decreases from $r_0$ down
to zero \cite{Haran}.

In a $d$-wave superconductor without an admixture of
$s$-wave, i. e., at $r=0$, the normalized
specific heat jump decreases monotonously down
to zero as $T_c/T_{c0}$ is suppressed from 1 to 0. For an arbitrary function
$\phi({\bf n})$ obeying the condition $\langle\phi({\bf n})\rangle_{FS}=0$ we
have at $T_c/T_{c0}<<1$
\begin{equation}
\frac{\Delta C}{C_n(T_c)}=\frac{8\langle\phi^2({\bf n})\rangle^2_{FS}}
{3\langle\phi^4({\bf n})\rangle_{FS}-2\langle\phi^2({\bf n})\rangle^2_{FS}}
\gamma^2\left(\frac{T_c}{T_{c0}}\right)^2,
\label{DeltaC3}
\end{equation}
where $\gamma=e^C\approx 1.781$ and $C$ is the Euler constant. Note that
$\langle\phi^4({\bf n})\rangle_{FS}>\langle\phi^2({\bf n})\rangle^2_{FS}$
for any $\phi({\bf n})$, so the denominator in Eq. (\ref{DeltaC3}) is
always positive. For our choice of $\phi({\bf n})=r+\cos(2\varphi)$ one has
$\Delta C/C_n(T_c)=(16/5)\gamma^2(T_c/T_{c0})^2$ at $r=0$ and $T_c<<T_{c0}$.

The effect of nonmagnetic impurities on the specific heat jump in
superconductors with different anisotropy of the order parameter is
illustrated in Fig. 2.
The difference in the behavior of $\Delta C/C_n(T_c)$ at
$T_c\rightarrow 0$ in superconductors with $r=0$ and $r\neq 0$ stems from the
fact that, while $T_c$ of a $d$-wave superconductor vanishes at a finite
value of $\rho_p=1/4\gamma\approx 0.140$ (i. e., at a finite
concentration of nonmagnetic impurities), the value of $T_c$ in a
superconductor with the nonzero Fermi surface average of $\Delta ({\bf p})$
asymptotically goes to zero as $\rho_p$ increases at $\rho_s=0$, see
Refs. \cite{Openov,Abrikosov1}.

\vskip 6mm

\centerline{\bf B. Pure spin-flip disorder}

\vskip 2mm

Although pure spin-flip scattering never happens in real materials since
magnetic impurities give rise to not only spin-flip scattering but to
potential scattering as well, we nevertheless (partly for pedagogical
purposes) consider briefly the limiting case that a superconductor contains
spin-flip scatterers only, i. e., $\rho_p= 0$ and $\rho=\rho_s/2$.
It is straightforward to show that in this case there
are no {\it qualitative} differences among the curves of $\Delta C/C_n(T_c)$
versus $T_c/T_{c0}$ for different values of $r$. From Eqs. (\ref{DeltaC1})
and (\ref{df/dDelta^2}) one finds that the normalized
specific heat jump decreases
monotonously down to zero with decreasing $T_c$, no matter what the symmetry
and the degree of anisotropy of $\Delta(\bf p)$ are, see Fig. 3. The physical
reason is that spin-flip scatterers, contrary to potential ones, are pair
breakers in both $d$-wave and $s$-wave superconductors \cite{Openov}. Two
quantitative differences between $d$-wave and $s$-wave symmetries
are (i) the different values of $\Delta C/C_n(T_c)$
at $T_c=T_{c0}$ and (ii) the different values of the coefficient
$f[\phi({\bf n})]$ in the dependence
$\Delta C/C_n(T_c)=f[\phi({\bf n})](T_c/T_{c0})^2$ at $T_c/T_{c0}<<1$.
For a $d$-wave superconductor, the value of $f[\phi({\bf n})]$
coincides with that in the case of nonmagnetic disorder, see
Eq. (\ref{DeltaC3}), while $f=8\gamma^2$ for an isotropic $s$-wave
superconductor, in agreement with Ref. \cite{Abrikosov}.

\vskip 6mm

\centerline{\bf C. Combined nonmagnetic and magnetic disorder}

\vskip 2mm

\centerline{1. \it Constant concentration of spin-flip scatterers}

\vskip 2mm

Now we turn to a general case that there are both nonmagnetic and magnetic
impurities in a superconductor. To begin with, we consider a situation when
purely nonmagnetic impurities are added to a superconductor that already
contains a small quantity of magnetic impurities and, as a consequence,
initially has the critical temperature $T_{c0}^{\prime}$ lower than
the value of $T_{c0}$ in the absence of any impurities. The higher is the
concentration of magnetic impurities, $c_m$, the greater is the value of
$\delta T_{co}/T_{co}$, where $\delta T_{co}=T_{c0}-T_{c0}^{\prime}$. We
assume that the value of $c_m$ remains unchanged upon increase in the
concentration of nonmagnetic impurities and corresponding decrease of $T_c$,
i. e., $\rho_{s0}=1/2\pi\tau_s T_{c0}=$ const. Since our prime interest here
is with the case $\delta T_{co}/T_{co}<<1$, we plot $\Delta C/C_n(T_c)$
versus $T_c/T_{c0}$ rather than $T_c/T_{c0}^{\prime}$.

In a $d$-wave superconductor, i. e., at $r=0$, the dependence of
$\Delta C/C_n(T_c)$ on $T_c/T_{c0}$ for any value of $\delta T_{co}/T_{co}$
is the same as in the absence of spin-flip scattering, see Fig. 2. This is
because at $\langle\phi({\bf n})\rangle_{FS}=0$ both $T_c$ and $\Delta C$ are
functions of the total scattering rate $\rho=\rho_p+\rho_s/2$ only,
irrespective of the scatterers' type, see Eqs. (\ref{Tc}), (\ref{DeltaC1}),
and (\ref{df/dDelta^2}).
Fig. 4 shows the curves of $\Delta C/C_n(T_c)$ versus
$T_c/T_{c0}$ for different values of $r\neq 0$ and $\delta T_{c0}/T_{c0}$. At
low $r=0.2$, i. e., in a $(d+s)$-wave
superconductor with a small admixture of
$s$-wave, or in a strongly anisotropic $s$-wave superconductor, the presence
of finite, though very small ($\delta T_{c0}/T_{c0}<0.01$) amount of
spin-flip scatterers in the sample results in a drastic change in the
dependence of $\Delta C/C_n(T_c)$ on $T_c/T_{c0}$. After passing through a
minimum, $\Delta C/C_n(T_c)$ does not increase up to $12/7\zeta(3)$, as in
the absence of magnetic impurities, but reaches the maximum and then
decreases again down to zero as $T_c\rightarrow 0$, see Fig. 4. It is
straightforward to show from Eqs. (\ref{Tc}), (\ref{DeltaC1}), and
(\ref{df/dDelta^2}) that
\begin{equation}
\frac{\Delta C}{C_n(T_c)}=\frac{1}{2\rho_{s0}^2}%
\left(\frac{T_c}{T_{c0}}\right)^2=
\frac{\pi^4}{32}\left[1+\frac{\langle\phi({\bf n})\rangle_{FS}^2}%
{\langle\phi^2({\bf n})\rangle_{FS}}\right]^2%
\left(\frac{T_c}{\delta T_{c0}}\right)^2
\label{DeltaC4}
\end{equation}
at $T_c<<\delta T_{c0}$. As the concentration of spin-flip scatterers (and,
hence, the value of $\delta T_{c0}/T_{c0}$) increases, the maximum of
$\Delta C/C_n(T_c)$ decreases in height and gradually disappears. Starting
with $\delta T_{c0}/T_{c0}\approx 0.02$ there are neither minimum nor
maximum of $\Delta C/C_n(T_c)$ versus $T_c/T_{c0}$ curve. We note that for
$T_{c0}\approx 100$ K such values of $\delta T_{c0}/T_{c0}$ correspond to
commonly observed small (several Kelvins) variations of $T_{c0}$ between the
samples obtained under slightly different conditions that can reflect the
different amount of magnetic impurities in the samples.

As $r$ increases, i. e., as the anisotropy of $\Delta({\bf p})$
becomes weaker, the tendency in the change of the specific heat jump upon
increase in the initial concentration of magnetic impurities is qualitatively
conserved, see Fig. 4. Quantitatively, an increase in $r$ results in the
increase in the value of $\delta T_{c0}/T_{c0}$ above which
$\Delta C/C_n(T_c)$ becomes a featureless monotonous function of
$T_c/T_{c0}$. In particular, both the minimum and the
maximum of $\Delta C/C_n(T_c)$
versus $T_c/T_{c0}$ curve disappear at $\delta T_{c0}/T_{c0}\approx$
0.1, 0.25, and 0.4 for $r$ = 0.6, 1, and 2, respectively. So, the sensitivity
of the specific heat jump to magnetic impurities is higher in strongly
anisotropic superconductors with small but nonzero values of $r$.

\vskip 6mm

\centerline{2. \it Constant ratio of spin-flip to potential scattering rates}

\vskip 2mm

Let us now consider the case that
the relative contribution from spin-flip scattering to the
total scattering rate, $\rho=\rho_p+\rho_s/2$, remains constant upon
disordering, i. e., the value of the dimensionless coefficient
\begin{equation}
\alpha=\frac{\tau_s^{-1}}{\tau_p^{-1}+\tau_s^{-1}} =
\frac{|u_m^{ex}|^2}{(c_n/c_m)|u_n|^2+|u_m^{pot}|^2+|u_m^{ex}|^2} =
\frac{\rho_s/2}{\rho_p+\rho_s/2} ,
\label{alpha}
\end{equation}
see Eq. (\ref{tau}), does not change upon addition of magnetic
(and, in general, nonmagnetic) impurities. This holds, first, if a
superconductor is doped by magnetic impurities only (i. e.,
$c_n=0$) and, second, if the ratio of nonmagnetic to magnetic
impurity concentrations, $c_n/c_m$, remains unchanged, see Eq.
(\ref{alpha}). The latter is a reasonable approximation for doping
by given chemical elements or irradiation by a given type of
particles, at least at relatively low (but sufficient to destroy
the superconductivity) doping levels or radiation doses.

Thus for a given degree of $\Delta({\bf p})$ anisotropy (i. e., in our model,
for a given value of $r$), the dependence of $\Delta C/C_n(T_c)$ on
$T_c/T_{c0}$ is governed by the value of material-dependent and
disorder-dependent coefficient $\alpha$. The greater is the relative
contribution from exchange scattering by magnetic impurities to the total
scattering rate, the higher is the value of $\alpha$. In general, $\alpha$
ranges from 0 in the absence of exchange scattering to 1 in the absence of
potential scattering. Note, however, that since there always exist two
channels of carrier scattering by magnetic impurities (potential and
spin-flip ones), see Eq. (\ref{tau}), the value of $\alpha$ is less than
unity even at $c_n=0$. Below we consider the case $\alpha << 1$ that seems to
be relevant to the experimental situation.

It follows from Eqs. (\ref{Tc}), (\ref{DeltaC1}), and (\ref{df/dDelta^2})
that in a $d$-wave superconductor with $r=0$ the curves of
$\Delta C/C_n(T_c)$ versus $T_c/T_{c0}$ are the same for any value of
$\alpha$ in the whole range of $\alpha$, see Fig. 2 and Fig. 3.
Contrary, at $r\neq 0$ the specific heat jump appears to be extremely
sensitive to spin-flip scattering of charge carriers.  Fig. 5a
shows the dependencies of $\Delta C/C_n(T_c)$ on $T_c/T_{c0}$ in a strongly
anisotropic non $d$-wave superconductor
with $r=0.2$ for different values of $\alpha$.
One can see that increase in $\alpha$ results in a gradual disappearance of
the minimum (and maximum) of $\Delta C/C_n(T_c)$. In the presence of even a
minor spin-flip component in the scattering potential, $\alpha\approx 0.02$,
the normalized specific heat jump decreases monotonously as $T_c$ decreases
from $T_{c0}$ down to zero, and the curve of $\Delta C/C_n(T_c)$ versus
$T_c/T_{c0}$ looks like that in a d-wave superconductor. As the gap
anisotropy weakens (i. e., the value of $r$ increases) the "critical" value
of $\alpha$ above which the normalized specific heat jump starts to decrease
monotonously under disordering first increases up to $\approx$ 0.05 at
$r\approx 1$ and next decreases again, see Fig.5.

\vskip 6mm

\centerline{\bf D. Implications for the experiment}

\vskip 2mm

Numerous experiments on various superconductors, including borocarbides
Y$_{1-x}$R$_x$Ni$_2$B$_2$C (R = Gd, Dy, Ho, Er) \cite{El-Hagary}, organic
compound (TMTSF)$_2$ClO$_4$ \cite{Pesty}, U$_{1-x}$Th$_x$Be$_3$
\cite{Scheidt}, HTSCs YBa$_2$(Cu$_{1-x}$M$_x$)$_3$O$_{7-\delta}$ (M = Zn
\cite{Loram,Meingast,Shamoto,Kim}, Fe \cite{Meingast}, Ni \cite{Shamoto},
Cr \cite{Kim}) and La$_{1.85}$Sr$_{0.15}$Cu$_{1-y}$Zn$_y$O$_4$ \cite{Mirza},
{\it etc} have revealed that the value of $\Delta C/C_n(T_c)$ decreases
monotonously as $T_c$ is suppressed by impurities. To the best of our
knowledge, there were no experimental indications for the nonmonotonous
behaviour of $\Delta C/C_n(T_c)$ in disordered superconductors. Note,
however, that the chemical substitution results
not only in the suppression of $T_c$ and decrease of $\Delta C/C_n(T_c)$ but
also in a very strong broadening of the superconducting transition. As a
consequence, the specific heat anomaly is rapidly smeared out by the
disorder, so that the dependence of $\Delta C/C_n(T_c)$ on $T_c/T_{c0}$
can be determined more or less reliably, in the best case, at
$T_c/T_{c0}>0.3\div 0.4$ only. Meanwhile,
it follows from the results presented above that the value of
$T_c/T_{c0}$ below which $\Delta C/C_n(T_c)$ starts to increase under
disordering depends on the degree of $\Delta({\bf p})$ anisotropy and is very
small in strongly anisotropic superconductors.

Recently Zhao has fitted $\Delta({\bf p})$ to single-particle tunneling and
angle-resolved photoemission spectra of
YBa$_2$Cu$_3$O$_{7-\delta}$ \cite{Zhao}. To compare his fit with our model
form of $\Delta({\bf p})$, it is convenient to introduce the coefficient
$\chi=%
1-\langle\Delta({\bf p})\rangle^2_{FS}/\langle\Delta^2({\bf p})\rangle_{FS}$
as a measure of the degree of in-plane anisotropy of $\Delta({\bf p})$, where
$\langle ... \rangle_{FS}$ means the Fermi surface average. The range
$0\leq\chi\leq 1$ covers the cases of isotropic $s$-wave
($\Delta({\bf p})$=const, $\chi=0$), $d$-wave
($\langle\Delta({\bf p})\rangle_{FS}=0$, $\chi=1$), and mixed $(d+s)$-wave
or anisotropic $s$-wave ($0<\chi<1$) symmetries of $\Delta({\bf p})$. Making
use of the results presented in Ref.\cite{Zhao}, one has $\chi\approx$ 0.9
for YBa$_2$Cu$_3$O$_{7-\delta}$. Since $\chi=1/(1+2r^2)$ for
$\Delta({\bf p})=\Delta[r+\cos (2\varphi)]$, this value of $\chi$
corresponds to $r\approx$ 0.2. We note that the choice
of $\chi\approx 0.9$ allows for a quantitative explanation \cite{Openov3} of
the quasilinear decrease of $T_c$ in electron-irradiated
YBa$_2$Cu$_3$O$_{7-\delta}$ single crystals \cite{Rullier}. As follows from
Figs. 4 and 5, at $r=0.2$, the upturn of $\Delta C/C_n(T_c)$ takes place at
$T_c/T_{c0}\approx 0.1$, either in the absence or at a very low concentration
of spin-flip scatterers.

So, the necessary condition for the $\Delta C/C_n(T_c)$ upturn at low
$T_c/T_{c0}$ is, except for the non-pure $d$-wave symmetry of
$\Delta({\bf p})$, a relatively small contribution of spin-flip scattering to
the total scattering rate. Besides, a superconductor should be disordered
very uniformly so that the transition width $\Delta T_c$ remained lower than
$T_c$ down to as low as possible $T_c$ values, in order to preserve a clear
specific heat anomaly at $T_c$ and to make possible the experimental
determination of the $\Delta C/C_n(T_c)$ versus $T_c/T_{c0}$ curve in a wide
region of $T_c/T_{c0}$ values. In this respect, the irradiation-induced
disorder has advantages over the chemical substitution.
For example, in a recent paper \cite{Rullier},
Rullier-Albenque {\it et al.} reported the results of experimental studies of
$T_c$ degradation under electron irradiation of underdoped and optimally
doped YBa$_2$Cu$_3$O$_{7-\delta}$ single crystals.
The authors of Ref. \cite{Rullier} succeeded in creation of an
extremely uniform distribution of radiation defects over the sample, so that
the value of $\Delta T_c$ never exceeded 5 K. Moreover, the value of
$\Delta T_c$ did not increase monotonously with radiation dose but had a
maximum at $T_c/T_{c0}\approx 0.3$ and next decreased down to
$\Delta T_c < 1$ K at the highest dose for which the resistive
superconducting transition still was observed at $T_c\approx 1$ K.
According to the theoretical fit \cite{Openov3} to the experimental data
\cite{Rullier}, at $\chi=0.9$, the value of $\alpha=0.01\pm 0.01$ is low
enough for the upturn of $\Delta C/C_n(T_c)$ be observable at
$T_c/T_{c0}\approx 0.1$, see Fig.5. Hence, it is of great interest to study
the behavior of $\Delta C$ versus $T_c/T_{c0}$ in such samples.

Finally, we note that the experimentally observed nonuniversality of
$\Delta C/C_n(T_c)$ versus $T_c/T_{c0}$ curve that has been previously
ascribed to the carrier concentration effects \cite{Shamoto} may in fact be
(at least partly) due to different contributions of spin-flip scattering to
pair breaking in different superconducting materials and/or for different
doping elements. We note also that it would be very interesting to study
experimentally the behavior of $\Delta C/C_n(T_c)$ versus $T_c/T_{c0}$ in
non-cuprate superconductors with different degree of the superconducting gap
anisotropy and various ratios of spin-flip to potential scattering rates.

\vskip 6mm

\centerline{\bf V. CONCLUSIONS}

\vskip 2mm

We have shown that in a pure $d$-wave superconductor, the normalized specific
heat jump $\Delta C/C_n(T_c)$ decreases monotonously upon disordering by both
nonmagnetic and magnetic defects or impurities. So, in $d$-wave
superconductors, $\Delta C/C_n(T_c)$ is, to a first
approximation (keeping in mind the assumptions made), a {\it universal}
function of $T_c/T_{c0}$, i. e., it does not depend on the relative
contribution of spin-flip scattering to the total scattering rate and, hence,
on a specific type of defects and impurities. On the other hand, under
nonmagnetic disordering of a superconductor with a nonzero Fermi surface
average of the order parameter, $\Delta C/C_n(T_c)$ initially
decreases with decreasing $T_c$, passes through a minimum and then increases
again. The minimum at the curve of $\Delta C/C_n(T_c)$ versus $T_c/T_{c0}$
moves to higher values of $T_c/T_{c0}$ as the anisotropy of the order
parameter becomes weaker.

In disordered strongly anisotropic non $d$-wave superconductors,
$\Delta C/C_n(T_c)$ is extremely sensitive to spin-flip scattering of charge
carriers. At relatively weak spin-flip scattering, $\Delta C/C_n(T_c)$
becomes a featureless monotonous function of $T_c/T_{c0}$. So, the spin-flip
scattering of charge carriers removes the qualitative difference between the
dependencies of $\Delta C/C_n(T_c)$ versus $T_c/T_{c0}$ in disordered
$d$-wave and $(d+s)$-wave (or anisotropic $s$-wave) superconductors. Hence,
it would be
very difficult to discriminate between pure $d$-wave and non pure $d$-wave
$\Delta({\bf p})$ if the concentration of spin-flip scatterers in the sample
is higher than a certain critical value. So, to observe the nonmonotonous
dependence of $\Delta C/C_n(T_c)$ on $T_c/T_{c0}$ in anisotropic
superconductors, one should (i) make use of uniformly disordered samples with
a clearly pronounced specific heat anomaly at $T_c$ down to low $T_c/T_{c0}$
values and (ii) minimize the
concentration of the spin-flip scatterers in the sample.

\vskip 6mm

\centerline{\bf ACKNOWLEDGMENTS}

\vskip 2mm

Discussions with R. Kishore and I. A. Semenihin are greatly acknowledged.
The work was supported in part by the Russian Department of Industry,
Science, and Technology under Grant No 40.012.1.1.1357.

\newpage

\includegraphics[width=\hsize]{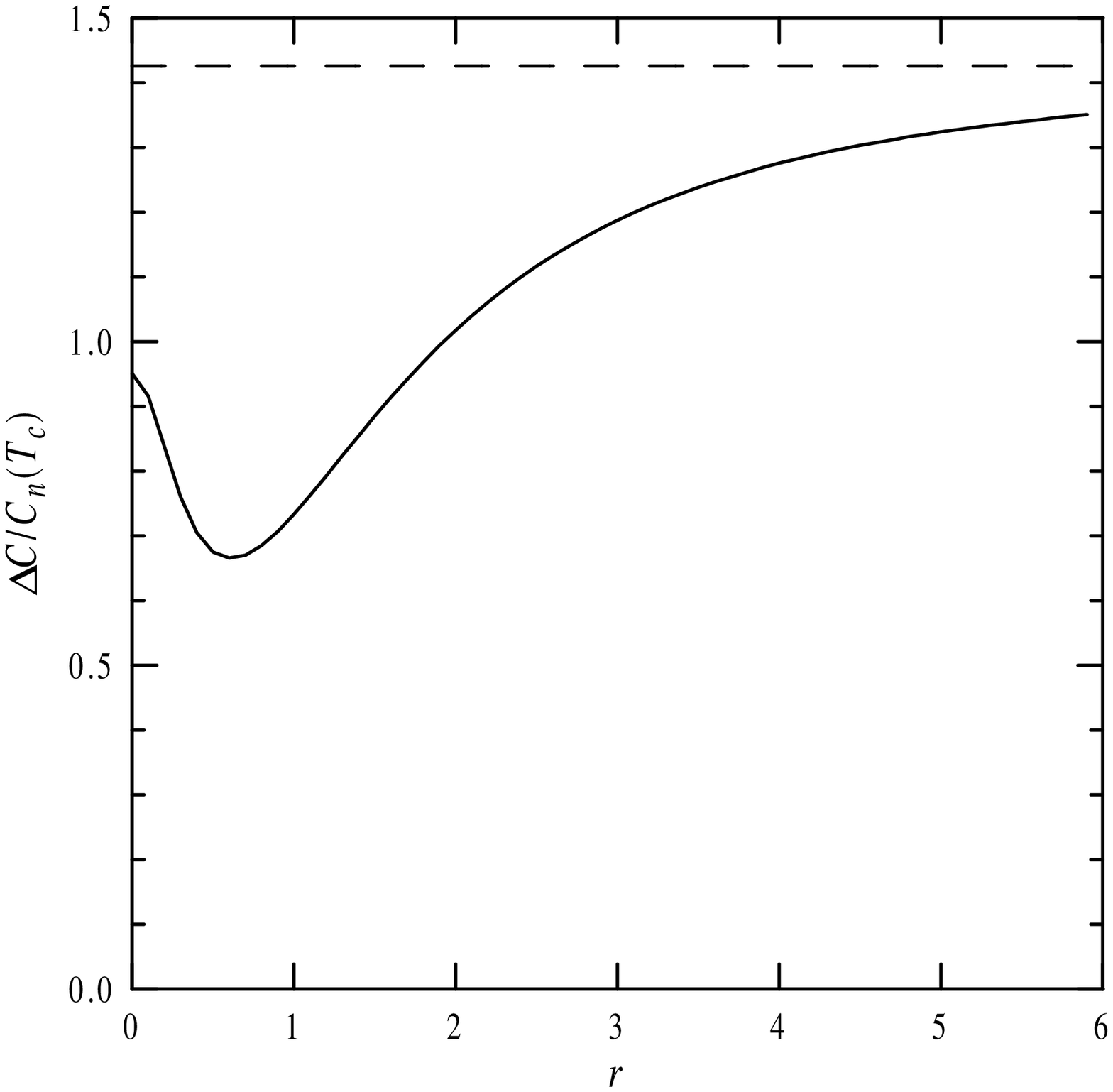}

Fig. 1. Specific heat jump $\Delta C$ normalized by the normal
state specific heat $C_n(T_{c0})$ versus the coefficient $r$ that
specifies the anisotropy of the superconducting order parameter
$\Delta({\bf p})=\Delta[r+\cos (2\varphi)]$, for a clean
superconductor without any impurities. Dashed line shows the value
of $\Delta C/C_n(T_{c0})=12/7\zeta(3)\approx$ 1.426 in an
isotropic s-wave superconductor.

\newpage

\includegraphics[width=\hsize]{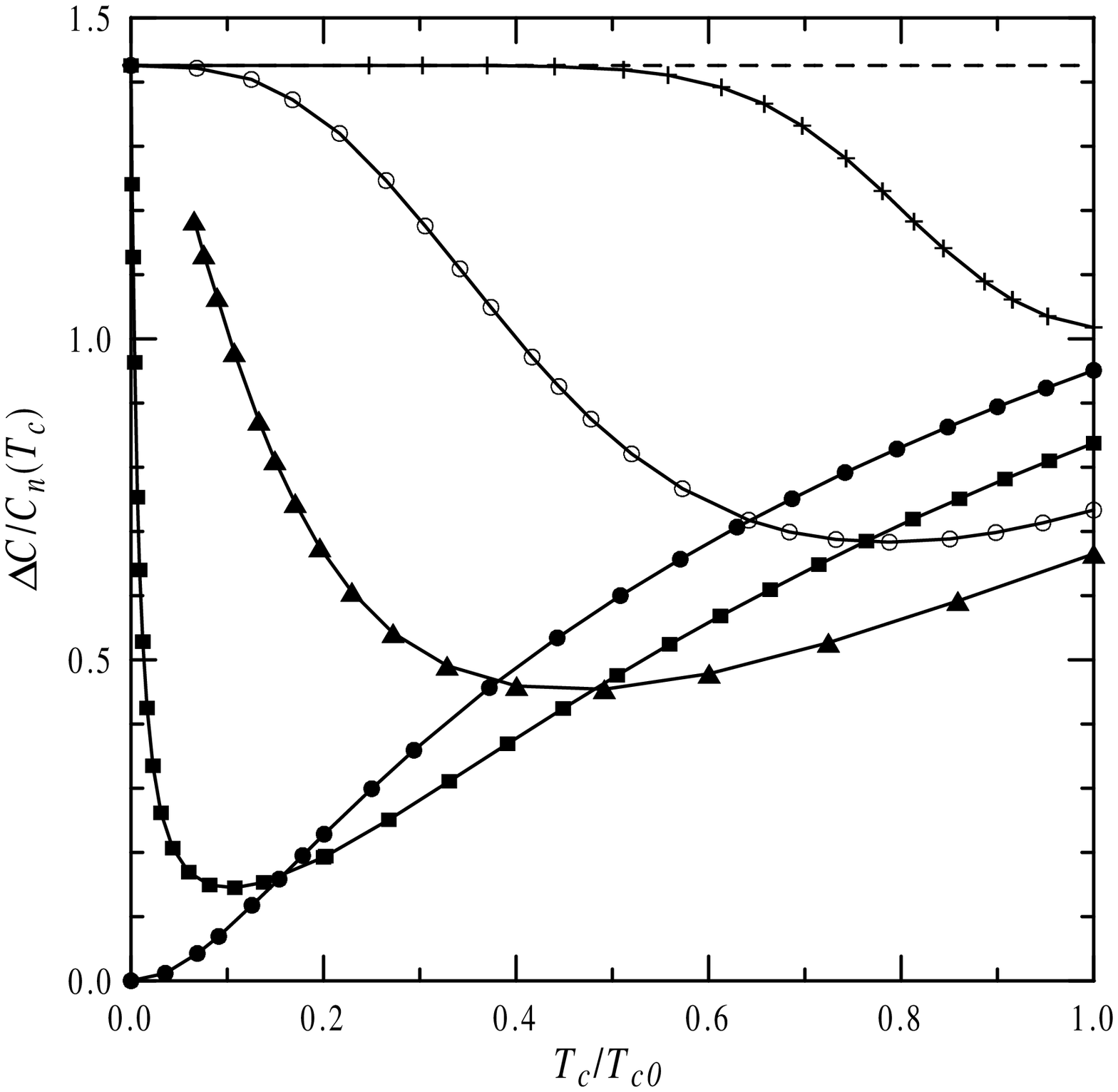}

Fig. 2. Specific heat jump $\Delta C$ normalized by the normal state specific
heat $C_n(T_c)$ versus the normalized critical temperature $T_c/T_{c0}$ for a
superconductor disordered by nonmagnetic impurities. The superconducting
order parameter is assumed to have the form
$\Delta({\bf p})=\Delta[r+\cos (2\varphi)]$, where $r=0$ (closed circles);
0.2 (squares); 0.6 (triangles); 1 (open circles); 2 (pluses). Solid lines are
guides for the eye. Dashed line corresponds to an isotropic s-wave
($r\rightarrow \infty, \Delta r\rightarrow $const).

\newpage

\includegraphics[width=\hsize]{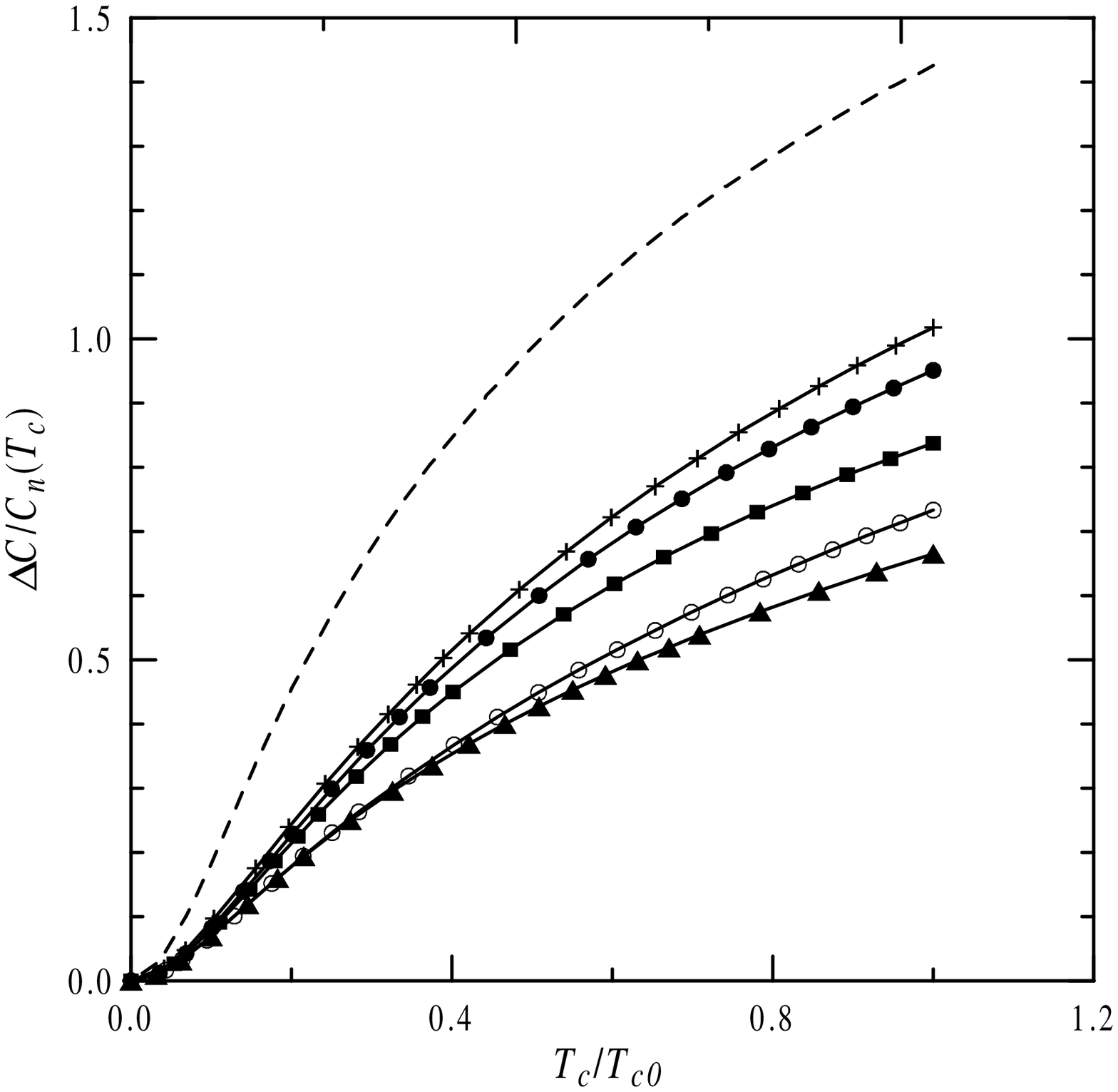}

Fig. 3. Same as in Fig. 2 for a superconductor disordered by spin-flip
scatterers only.

\newpage

\includegraphics[width=\hsize]{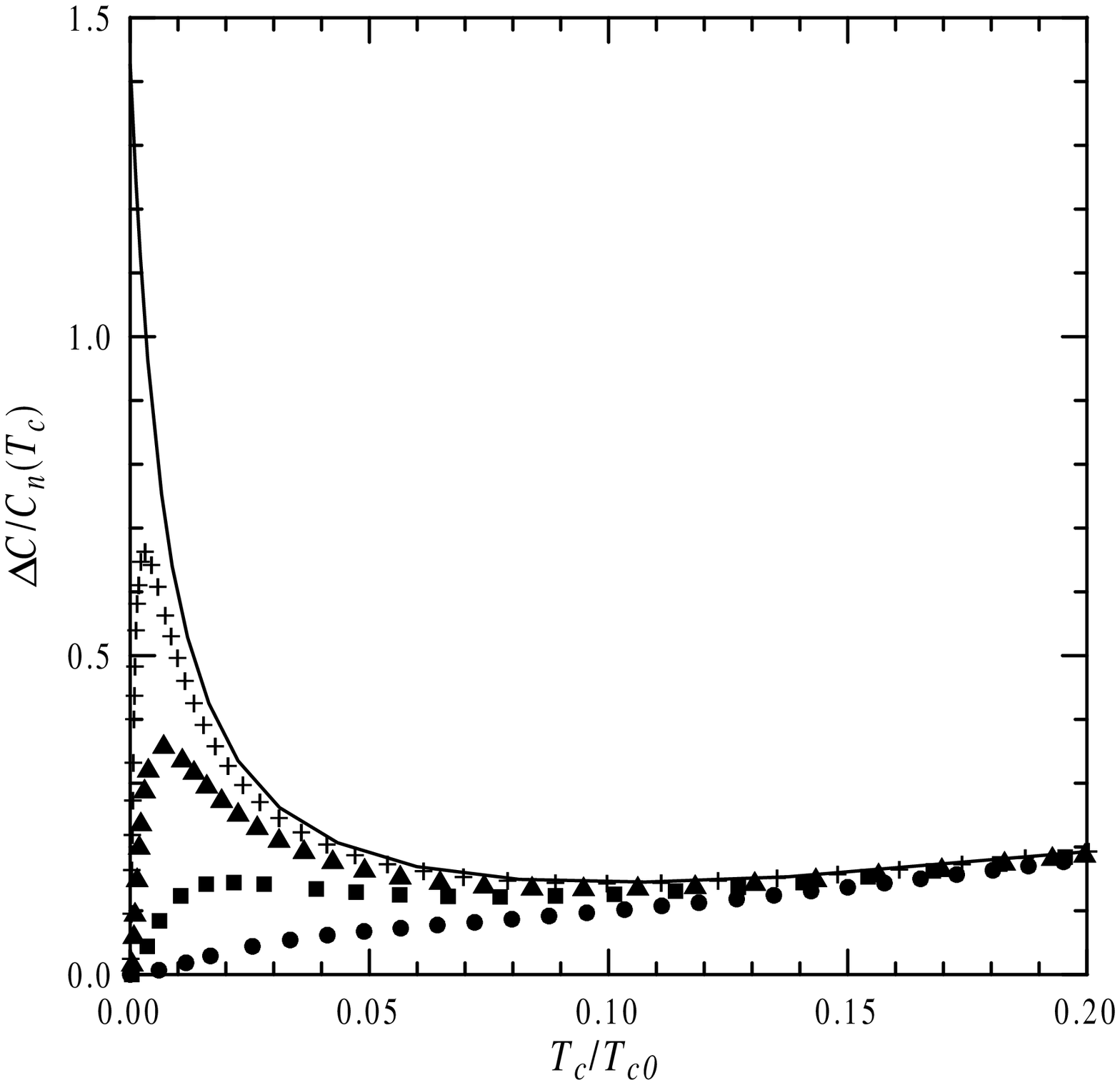}

Fig. 4a. Specific heat jump $\Delta C$ normalized by the normal
state specific heat $C_n(T_c)$ versus the normalized critical
temperature $T_c/T_{c0}$ for a superconductor with the order
parameter $\Delta({\bf p})=\Delta[r+\cos (2\varphi)]$. The
superconductor initially contains a small amount of spin-flip
scatterers and thus has the initial critical temperature
$T_{c0}^{\prime} < T_{c0}$. It is further disordered by
nonmagnetic impurities only, so that the concentration of
potential scatterers exceeds that of spin-flip scatterers.
$r=0.2$. $\delta T_{co}/T_{c0}=(T_{c0}-T_{c0}^{\prime})/T_{c0}$ =
0 (solid line), 0.001 (pluses), 0.003 (triangles), 0.01 (squares),
and 0.03 (circles). These values of $\delta T_{co}/T_{c0}$
correspond to the values of the spin-flip pair breaking rate
$\rho_{s0}=1/2\pi\tau_s T_{c0}=$ 0, 0.00038, 0.00113, 0.00376, and
0.01127 respectively. Note that at $T_c/T_{c0}>0.2$ the curves of
$\Delta C/C_n(T_c)$ versus $T_c/T_{c0}$ for different $\rho_{s0}$
(i. e., for different $\delta T_{co}$) almost coincide.

\newpage

\includegraphics[width=\hsize]{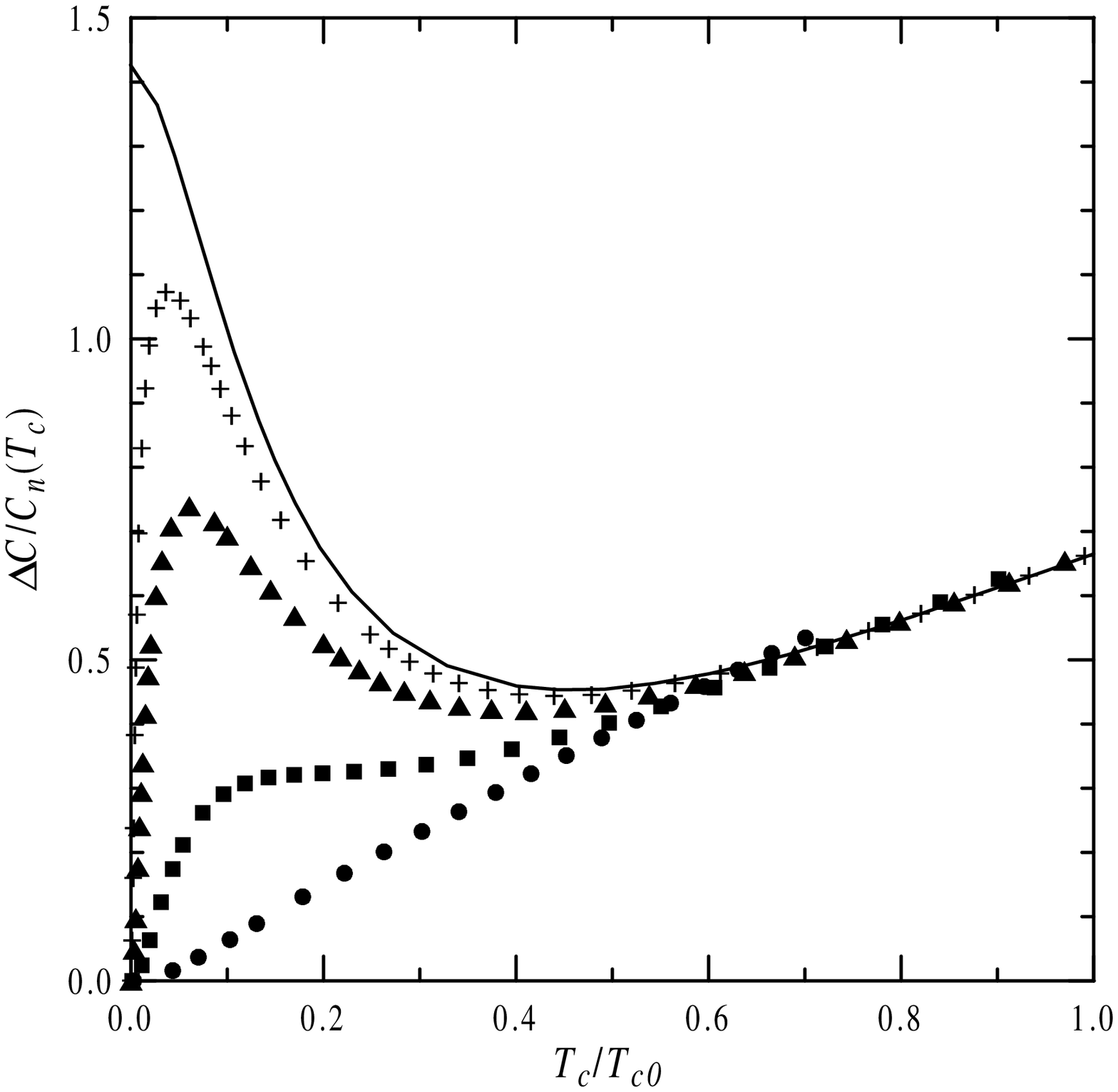}

Fig. 4b. Specific heat jump $\Delta C$ normalized by the normal
state specific heat $C_n(T_c)$ versus the normalized critical
temperature $T_c/T_{c0}$ for a superconductor with the order
parameter $\Delta({\bf p})=\Delta[r+\cos (2\varphi)]$. The
superconductor initially contains a small amount of spin-flip
scatterers and thus has the initial critical temperature
$T_{c0}^{\prime} < T_{c0}$. It is further disordered by
nonmagnetic impurities only, so that the concentration of
potential scatterers exceeds that of spin-flip scatterers.
$r=0.6$. $\delta T_{co}/T_{c0}=$ 0 (solid line), 0.01 (pluses),
0.03 (triangles), 0.1 (squares), and 0.3 (circles). [$\rho_{s0}=$
0, 0.00285, 0.00854, 0.02823, and 0.08232 respectively].

\newpage

\includegraphics[width=\hsize]{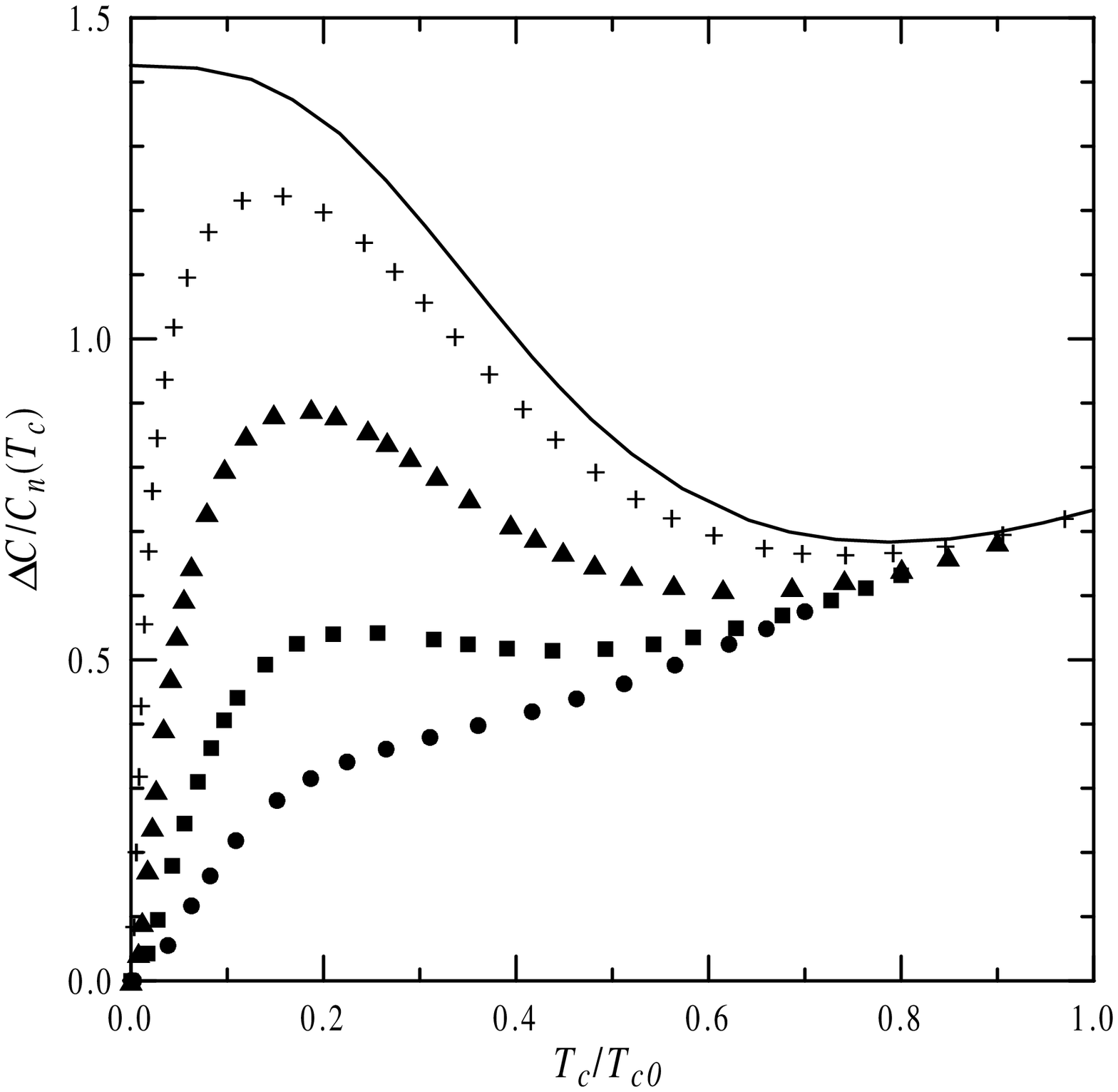}

Fig. 4c. Specific heat jump $\Delta C$ normalized by the normal
state specific heat $C_n(T_c)$ versus the normalized critical
temperature $T_c/T_{c0}$ for a superconductor with the order
parameter $\Delta({\bf p})=\Delta[r+\cos (2\varphi)]$. The
superconductor initially contains a small amount of spin-flip
scatterers and thus has the initial critical temperature
$T_{c0}^{\prime} < T_{c0}$. It is further disordered by
nonmagnetic impurities only, so that the concentration of
potential scatterers exceeds that of spin-flip scatterers. $r=1$.
$\delta T_{co}/T_{c0}=$ 0 (solid line), 0.03 (pluses), 0.1
(triangles), 0.2 (squares), and 0.3 (circles). [$\rho_{s0}=$ 0,
0.00727, 0.02400, 0.04728, and 0.06975 respectively].

\newpage

\includegraphics[width=\hsize]{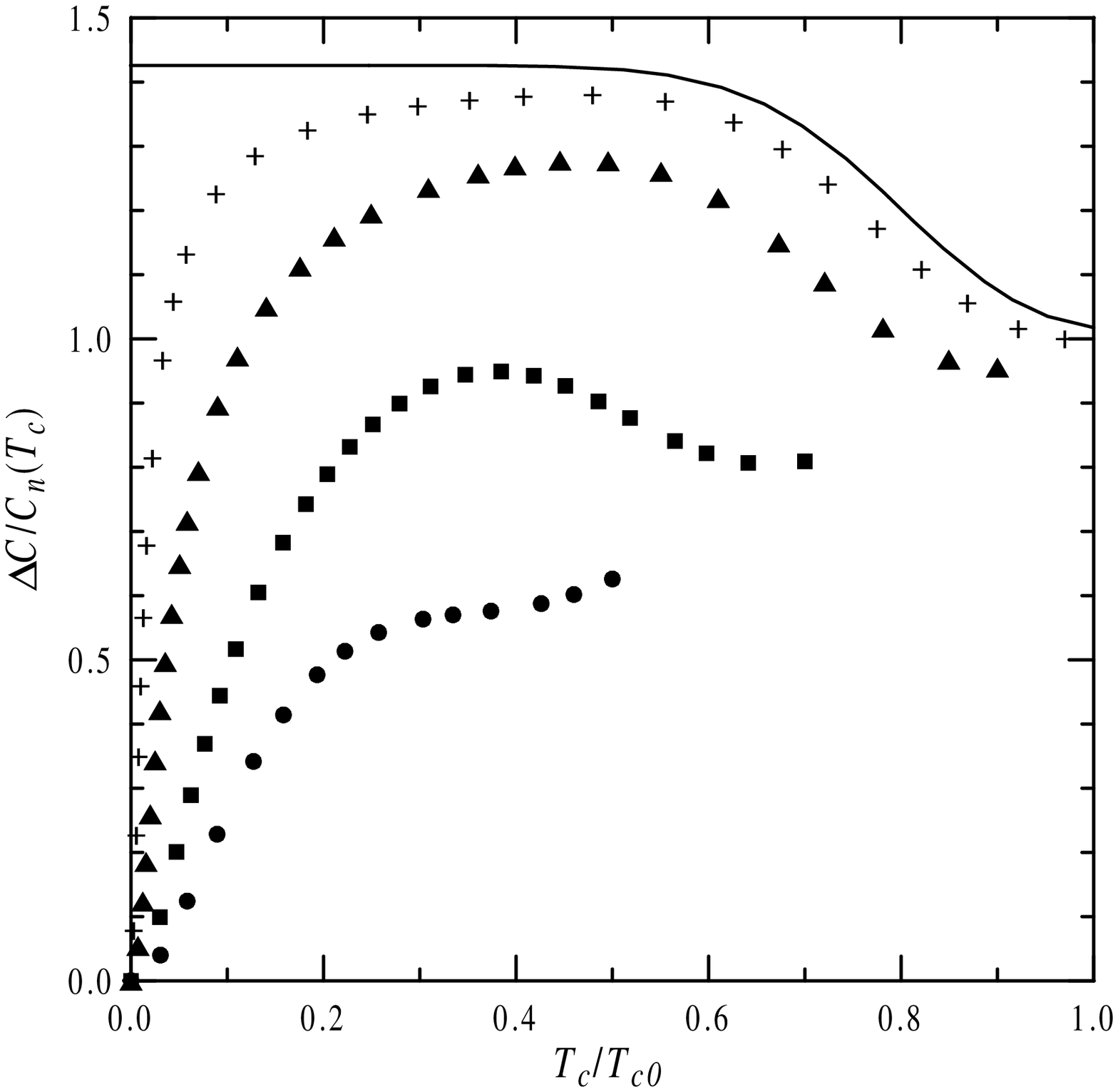}

Fig. 4d. Specific heat jump $\Delta C$ normalized by the normal
state specific heat $C_n(T_c)$ versus the normalized critical
temperature $T_c/T_{c0}$ for a superconductor with the order
parameter $\Delta({\bf p})=\Delta[r+\cos (2\varphi)]$. The
superconductor initially contains a small amount of spin-flip
scatterers and thus has the initial critical temperature
$T_{c0}^{\prime} < T_{c0}$. It is further disordered by
nonmagnetic impurities only, so that the concentration of
potential scatterers exceeds that of spin-flip scatterers. (d)
$r=2$. $\delta T_{co}/T_{c0}=$ 0 (solid line), 0.03 (pluses), 0.1
(triangles), 0.3 (squares), and 0.5 (circles). [$\rho_{s0}=$ 0,
0.00641, 0.02114, 0.06116, and 0.09731 respectively].

\newpage

\includegraphics[width=\hsize]{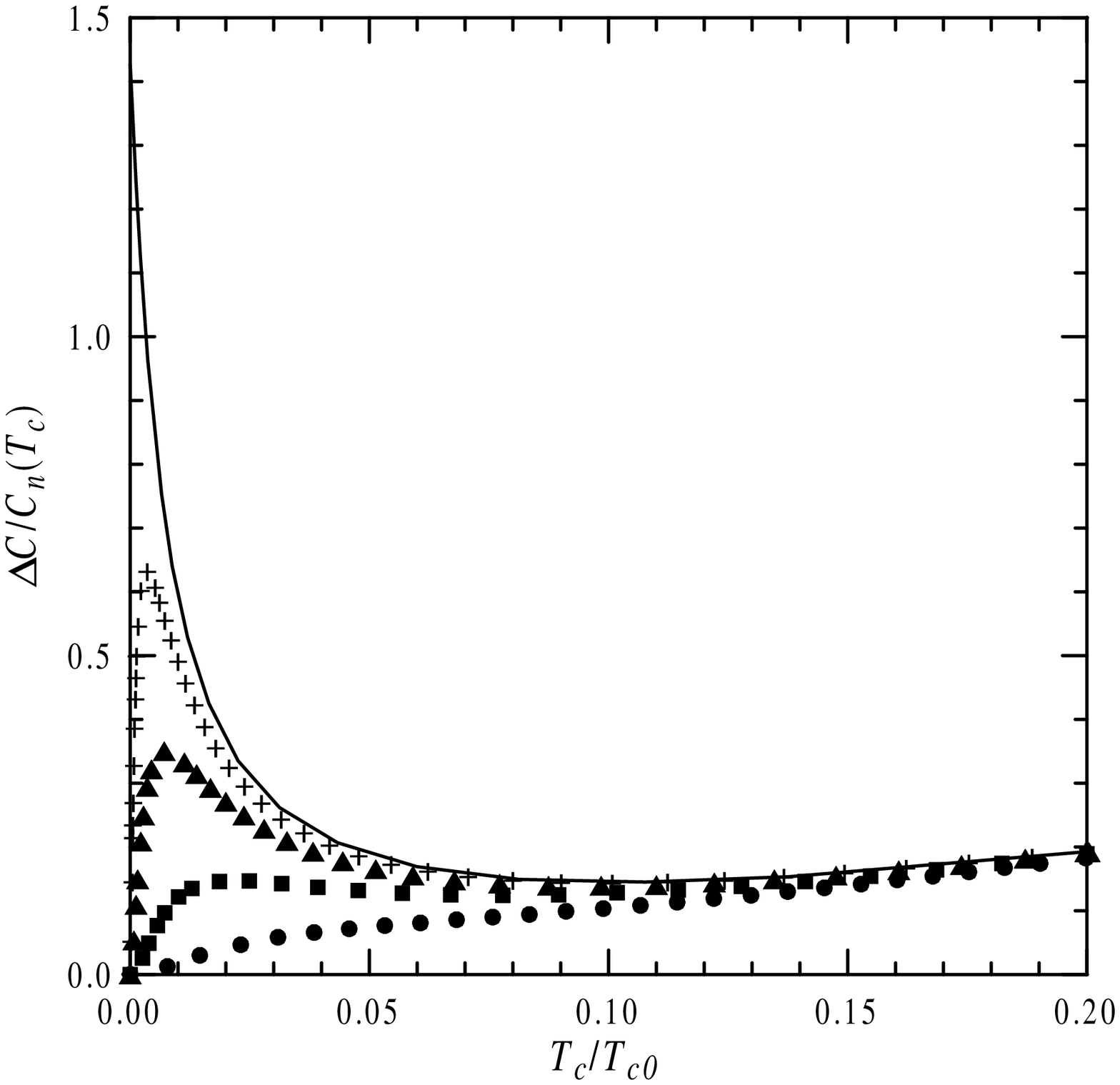}

Fig. 5a. Same as in Fig. 4 for a superconductor with a constant
ratio $\alpha$ of spin-flip to potential scattering rates.
$r=0.2$. $\alpha$ = 0 (solid line), 0.001 (pluses), 0.003
(triangles), 0.01 (squares), and 0.03 (circles).  Note that at
$T_c/T_{c0}>0.2$ the curves of $\Delta C/C_n(T_c)$ versus
$T_c/T_{c0}$ for different $\alpha$ almost coincide.

\newpage

\includegraphics[width=\hsize]{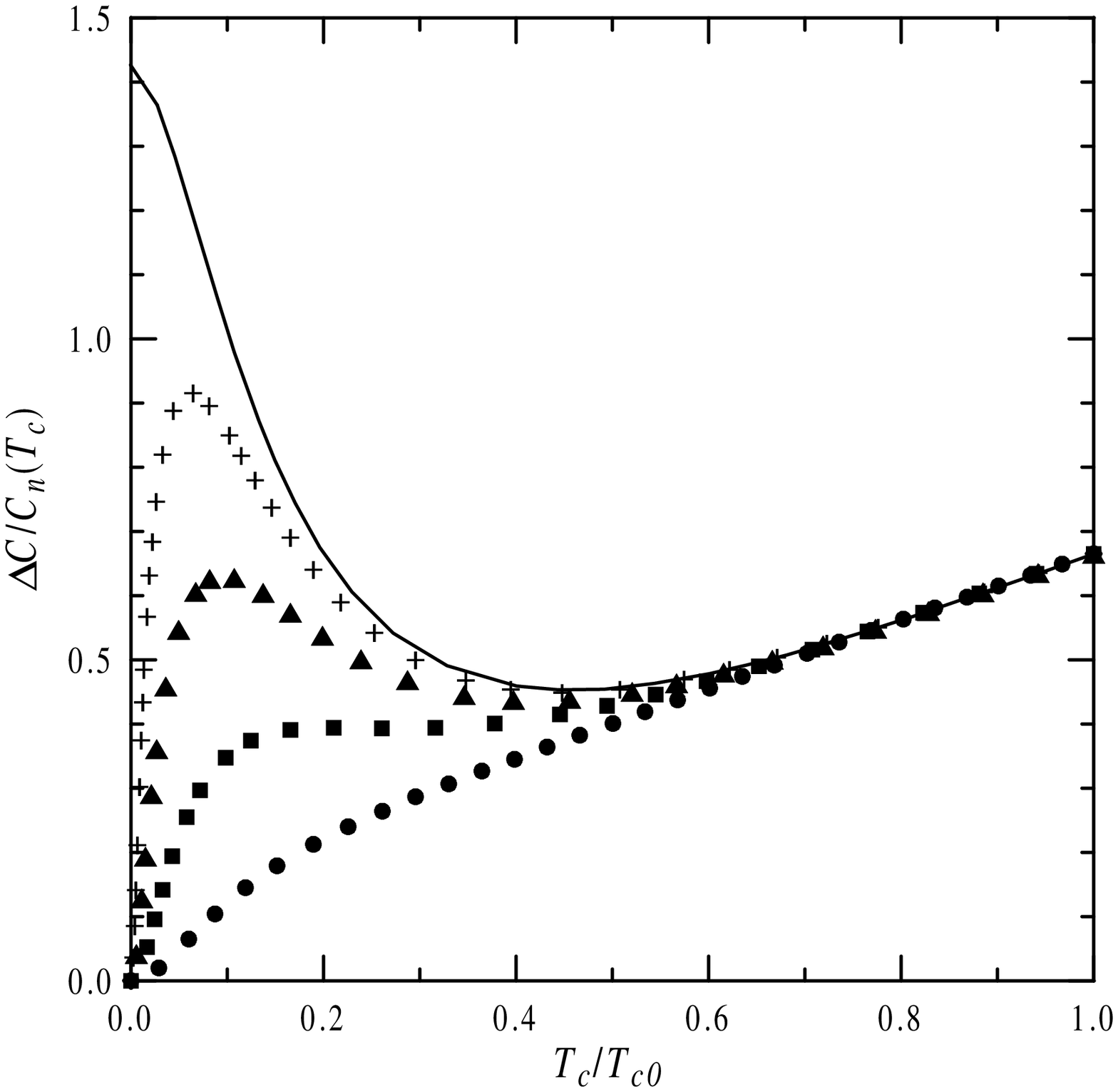}

Fig. 5b. Same as in Fig. 4 for a superconductor with a constant
ratio $\alpha$ of spin-flip to potential scattering rates.
$r=0.6$. $\alpha$ = 0 (solid line), 0.003 (pluses), 0.01
(triangles), 0.03 (squares), and 0.1 (circles).

\newpage

\includegraphics[width=\hsize]{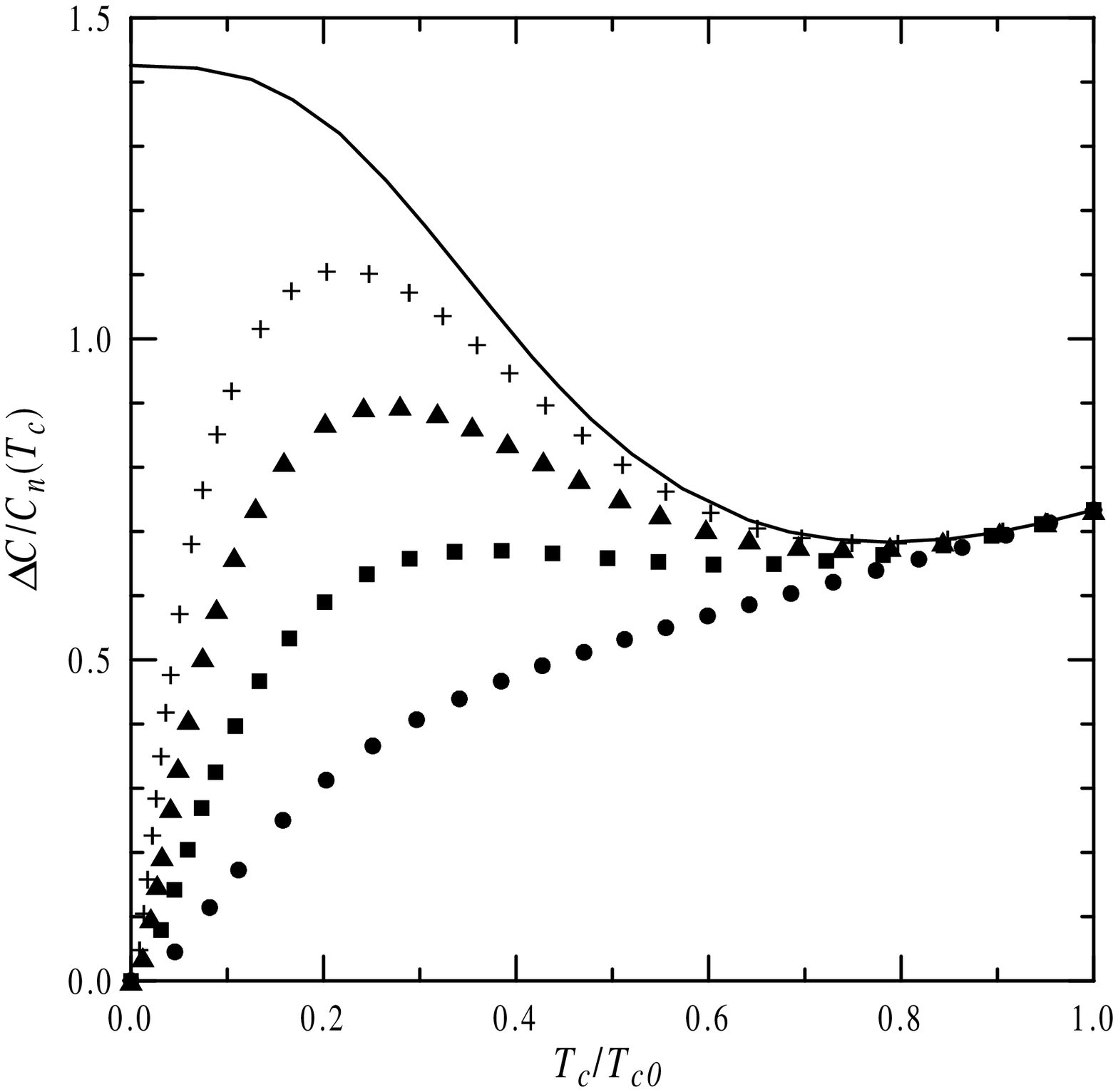}

Fig. 5c. Same as in Fig. 4 for a superconductor with a constant
ratio $\alpha$ of spin-flip to potential scattering rates. $r=1$.
$\alpha$ = 0 (solid line), 0.003 (pluses), 0.01 (triangles), 0.03
(squares), and 0.1 (circles).

\newpage

\includegraphics[width=\hsize]{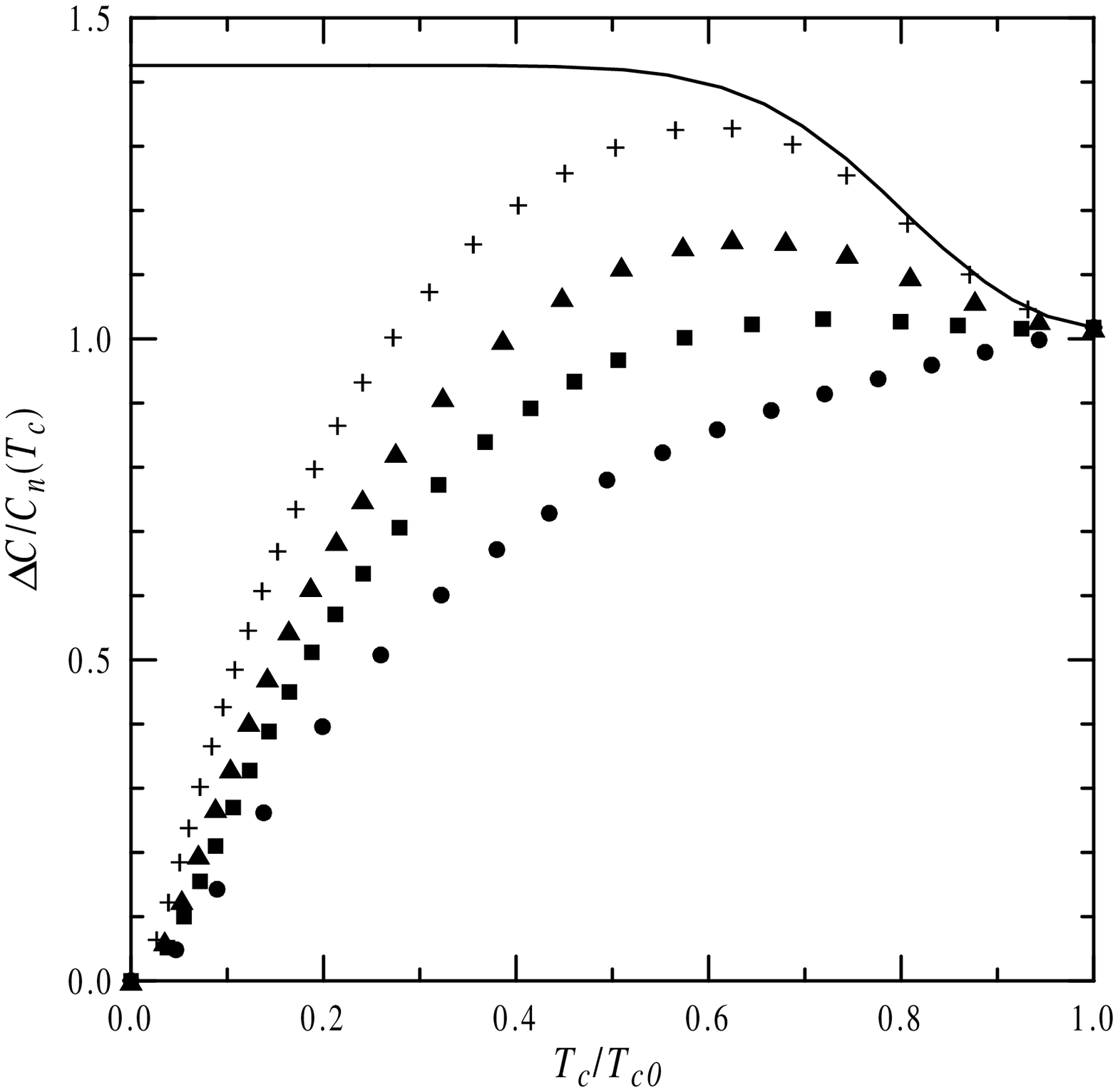}

Fig. 5d. Same as in Fig. 4 for a superconductor with a constant
ratio $\alpha$ of spin-flip to potential scattering rates. $r=2$.
$\alpha$ = 0 (solid line), 0.001 (pluses), 0.01 (triangles), 0.03
(squares), and 0.1 (circles).

\end{document}